\documentclass{article}
\usepackage{spconf,amsmath,graphicx, float}

\title{Articulation GAN: Unsupervised modeling of articulatory learning}
\name{Ga\v{s}per Begu\v{s}$^{1*}$\thanks{Ga\v{s}per Begu\v{s} and Alan Zhou contributed equally to this work.  \textit{Corresponding author: Ga\v{s}per Begu\v{s} (begus@berkeley.edu).}}, Alan Zhou$^{2*}$, Peter Wu$^{1\dagger}$, Gopala K. Anumanchipalli$^{1}$\sthanks{G.K.A.~and P.W.~are supported by NSF \#2106928.}}
\address{$^{1}$University of California, Berkeley, $^{2}$Johns Hopkins University
}
\usepackage{amsfonts}
\usepackage{amssymb}
\usepackage[safe]{tipa}
\usepackage[hidelinks]{hyperref}
\newcommand\rurl[1]{%
  \href{http://#1}{\nolinkurl{#1}}%
}
\begin{document}
\ninept
\maketitle
\begin{abstract}

Generative deep neural networks are widely used for speech synthesis, but most existing models directly generate waveforms or spectral outputs. Humans, however, produce speech by controlling articulators, which results in the production of speech sounds through physical properties of sound propagation. We introduce the Articulatory Generator to the Generative Adversarial Network paradigm, a new unsupervised generative model of speech production/synthesis. The Articulatory Generator more closely mimics human speech production by learning to generate articulatory representations (electromagnetic articulography or EMA) in a fully unsupervised manner. A separate pre-trained physical model (ema2wav) then transforms the generated EMA representations to speech waveforms, which get sent to the Discriminator for evaluation. Articulatory analysis suggests that the network learns to control articulators in a similar manner to humans during speech production. Acoustic analysis of the outputs suggests that the network learns to generate words that are both present and absent in the training distribution. We additionally discuss implications of articulatory representations for cognitive models of human language and speech technology in general.

\end{abstract}
\begin{keywords}
articulatory phonetics, unsupervised learning, electromagnetic articulography, deep generative learning
\end{keywords}
\section{Introduction}
\label{intro}

Humans produce spoken language with articulatory gestures \cite{stevens98}. Sounds of speech are generated by airflow from the lungs passing through articulators, which causes air pressure fluctuations that constitute sounds of speech. The main mechanism in speech production is thus control of the articulators and airflow  \cite{stevens98}. During language acquisition, children need to learn to control articulators and produce articulatory gestures such that the generated sounds correspond to the sounds of language they are exposed to. 

This learning is complicated by the fact that sound is an entirely different modality compared to articulatory gestures.  Children need to learn to control and move articulators from sound input without direct access to the articulatory data of their caregivers. While some articulators are visible (such as lips and tongue tip, jaw movement), many are not (vocal folds, tongue dorsum). There is debate on whether spoken language acquisition is fully unsupervised due to direct and indirect negative evidence  \cite{lust06}. Articulatory learning, however, is likely fully unsupervised. Caregivers ordinarily do not provide any explicit feedback about articulatory gestures to language-acquiring children.

Most models of human speech production output audio data of speech without articulatory representations. In actual speech,  however, humans control articulators and airflow, while  a separate physical process results in sounds of speech.

To build a more realistic model of human spoken language, we propose a new deep learning architecture within the GAN framework \cite{goodfellow14,radford15,donahue19,begus19,begusCiw,begus22Interspeech}. In our proposal, the decoder (synthesizer or the Generator network) learns to output approximates of human articulatory gestures while never accessing articulatory data. The generated articulatory gestures are represented with thirteen channels that match the twelve channels used to record human articulators during electromagnetic articulography (EMA) plus an additional channel for voicing. 
The generated articulatory movements are then passed through a separate \textit{physical model}  of sound generation that takes articulatory channels and converts them into waveforms. This physical model is taken from a pre-trained EMA-to-speech model (ema2wav) which transforms electromagnetic articulography into speech waveforms \cite{wu22}. This physical model component is a model of physical sound propagation and is cognitively irrelevant, which is why its weights are not updated during training.

Articulatory learning in this model needs to happen in a fully unsupervised manner. The Articulatory Generator needs to transform random noise in the latent space into the thirteen channels such that the independent pre-trained EMA-to-speech physical model will generate speech. The Discriminator receives waveform data synthesized based on the Articulatory Generator's generated channels. The Generator in our model never directly accesses articulatory data. Like humans, it needs to learn to control articulators without ever directly accessing  them (e.g.~vocal folds or tongue dorsum are never visible during speech acquisition). The only information available to humans during acquisition and our model during training is the auditory feedback from the perception component of speech that corresponds to the Discriminator network in our model.

\subsection{Prior work}

Speech synthesis from articulatory representations has recently been performed using deep neural networks \cite{bocquelet14,aryal16,chen21,georges22,georges2020,wu22}. The objective in most existing proposals, however, is to synthesize waveforms from articulatory representations in a supervised setting,  rather than a fully unsupervised generation of the articulatory representations themselves. \cite{mirrornet,mirrornet1} proposes an autoencoder model that learns to encode and decode between motor parameters and auditory representations in an unsupervised manner. However, this model trains both the encoding and decoding aspects of the model simultaneously, and focuses on the relationship between auditory representations and a motor latent space. By contrast, our GAN model is trained with a static pretrained articulatory model similar to how children learn to speak with a full set of articulators. In addition, rather than decoding back and forth between motor and auditory information, our model is able to generate articulatory parameters directly by sampling from a general-purpose latent space. To our knowledge, this paper presents the first architecture in which a generative model learns to produce unprompted articulatory gestures that result in speech in a fully unsupervised way.

Computational models of language almost always disregard the articulatory component. Currently, articulatory phonology is a proposal that comes closest to modeling linguistic representations from articulatory representations \cite{browman92,smith21emergent}, and phonological structure can be inferred from articulatory data \cite{baljekar15}. However, these models take articulatory gestures as a given (as measured on human subjects) and  do not model unsupervised learning and generation of articulatory gestures from auditory feedback.

\begin{figure*}
    \centering
    \includegraphics[width=.75\textwidth]{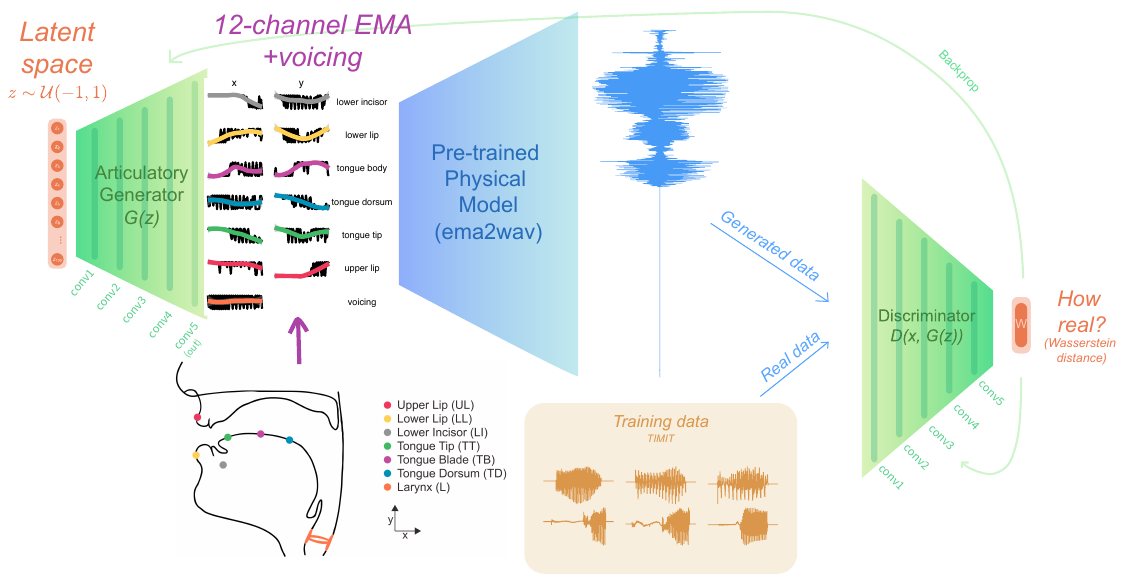}
    \caption{The architecture of the ArticulationGAN. The Articulatory Generator takes 100 latent variables $z$ and generates 12 EMA channels and the channel for voicing. The pre-trained physical model (ema2wav) takes the generated EMA and transforms them into waveforms.     }
    \label{fig:icassp}
\end{figure*}

A model of unsupervised articulatory learning is not only a more realistic representation of human speech, but is useful for conducting cognitive simulations that have the potential to reveal which properties of speech emerge because of articulatory factors and which properties are cognitively conditioned \cite{begusLanguage}. 
In engineering application, learning to generate plausible articulatory gestures with accompanied synthesized speech is useful for lip synchronization \cite{li21} (with potential applications in robotics or gaming industry). The modelling of articulatory information has also been identified as being useful in the detection of audio deepfakes \cite{blue22}. Generation of articulatory gestures is thus potentially useful for creating more realistic speech synthesis technologies, as well as providing another adversarial approach that deepfake detectors can use to improve their accuracy.

\section{The model}

Our articulatory model takes the architecture of WaveGAN \cite{donahue19}, and replaces its Generator with a combination of an Articulatory Generator and a physical model of articulation. The Articulatory Generator is a modification of the WaveGAN Generator that maps random noise to 13 channels of time-series data corresponding to articulatory representations and voicing. The physical model is a pretrained autoregressive encoder that maps the modified Generator's articulatory output into speech data. Note that the weights of the physical model are frozen during training: we constrain the problem so that the Articulatory Generator learns to produce articulatory movements that will result in realistic speech.

\subsection{Articulatory Generator}
The Articulatory Generator $G$ is adapted from the Generator network from WaveGAN \cite{donahue19}. It takes as input a latent noise vector $z$ and uses 5 layers one-dimensional transpose convolutions to upsample the noise into waveform data $G(z)$. Unlike WaveGAN, our Articulatory Generator generates 13 channels (one channel of voicing plus the x- and y-axis for 6 articulators). Due to the physical model's low sample rate, the dimensionality of each layer is also lower than in WaveGAN, with individual dimensionalities of $32 \times 512, 64 \times 512, 128 \times 256, 128 \times 256, 256 \times 13$, respectively.

\subsection{Physical Model}
We take the EMA-to-speech encoder trained on MNGU0 from \cite{wu22} to be a physical model of articulation $\mathcal{A}$. This autoregressive model takes as input 13 channels of time-series  and outputs a 16 kHz waveform corresponding to speech. Specifically, the 13 channels of articulatory features include one channel of voicing, plus the $x$ and $y$ coordinate positions each of the lower incisor, upper lip, lower lip, tongue tip, tongue body, and tongue dorsum.

\section{Training}

We train our model using the same WGAN-GP scheme \cite{gulrajani17}  as in \cite{donahue19}, except we replace the Generator's output with the output of our two-step articulatory inference:

$$
\max_D \min_G V(D, G) = \mathbb{E}_{x \sim P_x}[D(x)] - \mathbb{E}_{z \sim P_z}[D(\mathcal{A}(G(z)))]
$$

where $P_x$ and $P_z$ are the training and noise distributions, and the Discriminator $D$ is constrained to be 1-Lipshitz function.

We train the network on 8 words from TIMIT \cite{timit}: \textit{ask}, \textit{dark}, \textit{year}, \textit{water}, \textit{wash}, \textit{rag}, \textit{oily}, and \textit{greasy} for 354,200 training steps with a batch size of 8. These specific words were chosen because of their relatively equally frequent appearance in TIMIT.\footnote{Counts of each token as well as generated EMA and waveform data, annotations, and checkpoints are available at \rurl{doi.org/10.17605/OSF.IO/X37HA}. The code is available at \rurl{github.com/gbegus/articulationGAN}.} We limit the number of training words to facilitate learning as well as to mimic language acquisition more closely: productive vocabulary size is relatively small at the initial stages of language acquisition \cite{fenson94}.

The training data for the Generator is different from the MNGU0 data set \cite{richmond11} training dataset used in the EMA-to-speech physical model (which involves a single speaker of British English). This mimics human language acquisition, where children need to learn from multiple adults while having a single set of articulators. Our training is additionally complicated by MNGU0 and TIMIT involving speakers of different varieties of English (British vs~American).

We additionally train an unmodified WaveGAN network \cite{donahue19} on the same 8 words from TIMIT \cite{timit} as a baseline to compare against our articulatory model. This model was trained for 138,600 steps with a batch size of 32.

\section{Results}

\subsection{Performance}

To test how well the ArticulationGAN performs compared to WaveGAN, we randomly generated 200 outputs from the ArticulationGAN and WaveGAN models (400 total). A trained phonetician who is not a coauthor was hired to annotate and transcribe the outputs in order to avoid potential bias  and to account for noise in the outputs. The outputs were annotated as (i) intelligible words of English, (ii) intelligible sequences of sounds that are not words of English, and (iii) unintelligible outputs. The results are given in Figure \ref{tab:intelligible}.  Intelligible outputs include all sounds that were transcribable by the trained phonetician; the proportion of comfortably intelligible outputs is likely lower. The WaveGAN model performs slightly better on the intelligibility task (87\% vs.~72\%), but the ArticulationGAN outputs a higher proportion of intelligible outputs (words and non-words) that are not part of training data (innovative outputs).

\begin{table}
    \centering
    \begin{tabular}{lcc|c}
   \hline
      Model       & Intelligible & Unintelligible & Innovative \\  \hline
  WaveGAN       &174 (87\%) & 26 (13\%) &87 (50\%)\\
  ArticulationGAN      & 143 (72\%) &57 (29\%)&110 (77\%)\\
    \hline
    \end{tabular}
    \caption{Counts of annotated outputs in WaveGAN and ArticulationGAN architectures. The 200 annotated words per model are divided into intelligible and unintelligible outputs. The Innovative column indicates those intelligible outputs (words and non-words) that are not part of training data. 33 (17\%) outputs are training data words in ArticulationGAN (compared to 87 or 44\% in WaveGAN).}
    \label{tab:intelligible}
\end{table}

The weaker intelligibility of ArticulationGAN is likely due to its more difficult training objective. The WaveGAN Generator only needs to produce outputs that are themselves similar to the training distribution drawn from TIMIT. On the other hand, the Articulatory Generator must produce outputs that approximate the TIMIT data after being passed through a fixed articulatory model that has been trained on MNGU0 data. As previously mentioned, TIMIT data is drawn from 630 speakers across eight dialects of American English \cite{timit}, while MNGU0 data is drawn from a single speaker of British English \cite{richmond11}.

\begin{figure}[t]
    \centering
    \includegraphics[width=.35\textwidth]{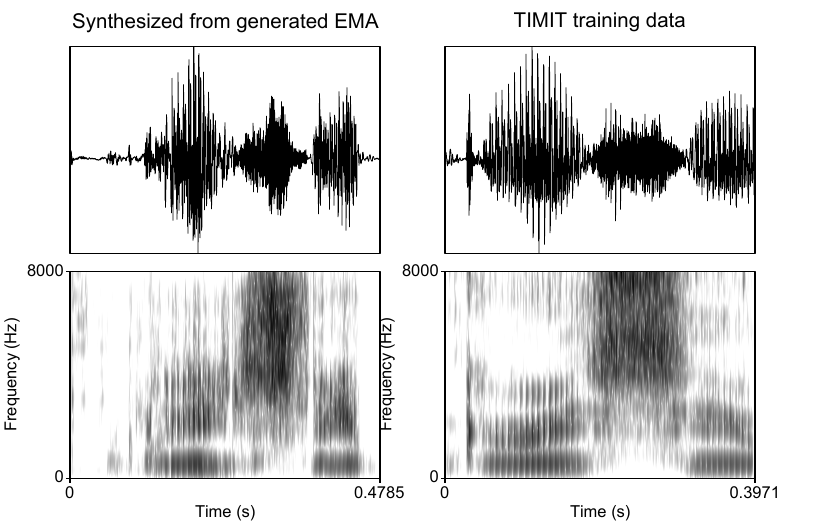}
    
    \caption{Generated output \textit{greasy} and its corresponding (TIMIT) datapoint used during training.}
    \label{fig:greasy}
\end{figure}

Nevertheless, ArticulationGAN produces a higher amount of innovative data compared to both the TIMIT and MNGU0 datasets. The results suggest that the ArticulationGAN not only learns words that are represented in both TIMIT training data and MNGU0 dataset (e.g.~\textit{wash}), but also words that are absent from the MNGU0 dataset and the TIMIT training dataset. For example, the ArticulationGAN generated outputs that were transcribed as \textit{wash} [\textipa{"wOS}], \textit{fast} [\textipa{"f\ae st}], \textit{greasy} [\textipa{"g\textturnr isi}], and \textit{coffee} [\textipa{"kOfi}].  \textit{Wash} is part of TIMIT  and MNGU0 training data.  \textit{Fast} and \textit{coffee} are only  present in the MNGU0 data. \textit{Fast} is acoustically close to \textit{ask} in the training data. \textit{Coffee} is distant to its closest equivalent in the TIMIT training data (\textit{greasy}), but \textit{greasy} is absent from MNGU0. The \textit{ema2wav} model is never trained to generate \textit{greasy} from EMA, yet our ArticulationGAN generates several outputs that can be reliably transcribed as \textit{greasy} (Figure \ref{fig:greasy}).

We also observe overrepresentation of \textit{w-}initial words in the 200 outputs of the ArticulationGAN compared to TIMIT training data ($\text{OR}=1.53, p<0.01$), but not in  WaveGAN outputs ($\text{OR}=0.94, p=0.74$). Gestures for [w-] are easier to acquire compared to other initial consonants. It appears that ease of articulation of labial consonants plays a role in articulatory learning in our models (similar to language acquisition \cite{vihman91,mcleod18}).

\subsection{Analyzing generated gestures}

To analyze unsupervised learning of articulatory gestures in the ArticulationGAN model, we compare real (MNGU0) and generated (ArticulationGAN) EMA channels and corresponding acoustic outputs (waveforms). We analyze articulatory gestures in two generated outputs transcribed as  \textit{wash} [\textipa{"wOS}] and \textit{fast} [\textipa{"f\ae st}]. These words were chosen because \textit{wash} is present in both TIMIT training and MGNU0 data, while \textit{fast} is an innovative output. Because \textit{greasy} is fully absent from MNGU0 training data, we cannot compare generated and real EMA for this word.

\begin{figure}[t]
    \centering
    \includegraphics[width=.35\textwidth]{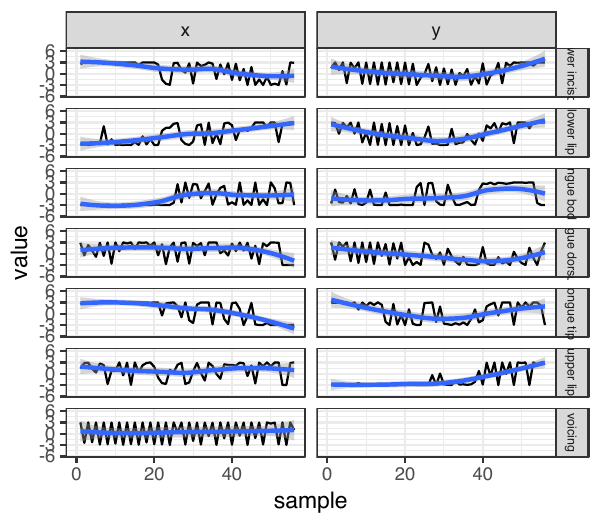}
    
    \caption{Generated EMA channels and voicing for \textit{wash} with LOESS smoothing.}
    \label{fig:LOESS}
\end{figure}

Figure \ref{fig:LOESS} illustrates the 12 channels plus voicing for \textit{wash}. We observe the network learns relatively stable articulatory targets, except during transitions between targets or when an articulator does not play an active role for a given phoneme sequence. For example, the x axis of tongue dorsum and the lower incisor position do not play a central role in the articulation of \textit{wash}, which is why this channel is relatively noisy in Figure \ref{fig:LOESS}.

\begin{figure}[t]
    \centering
    \includegraphics[width=.4\textwidth]{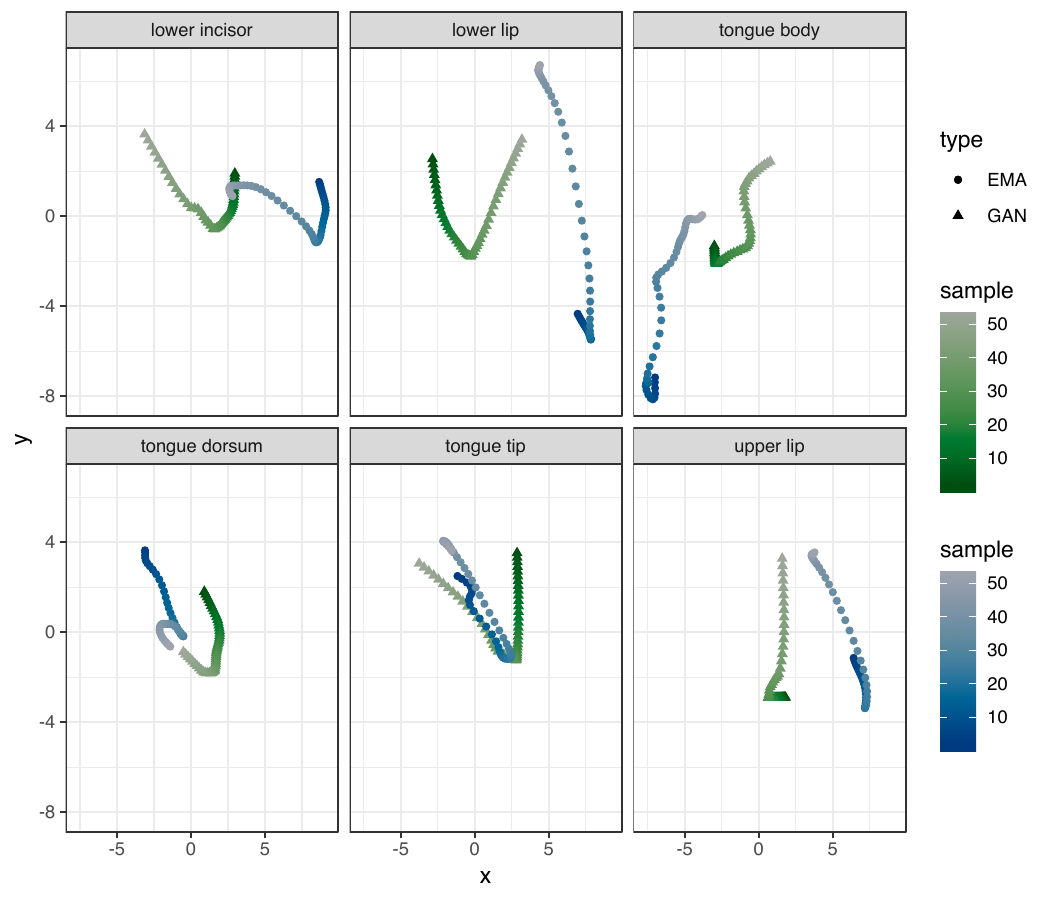}
    \includegraphics[width=.4\textwidth]{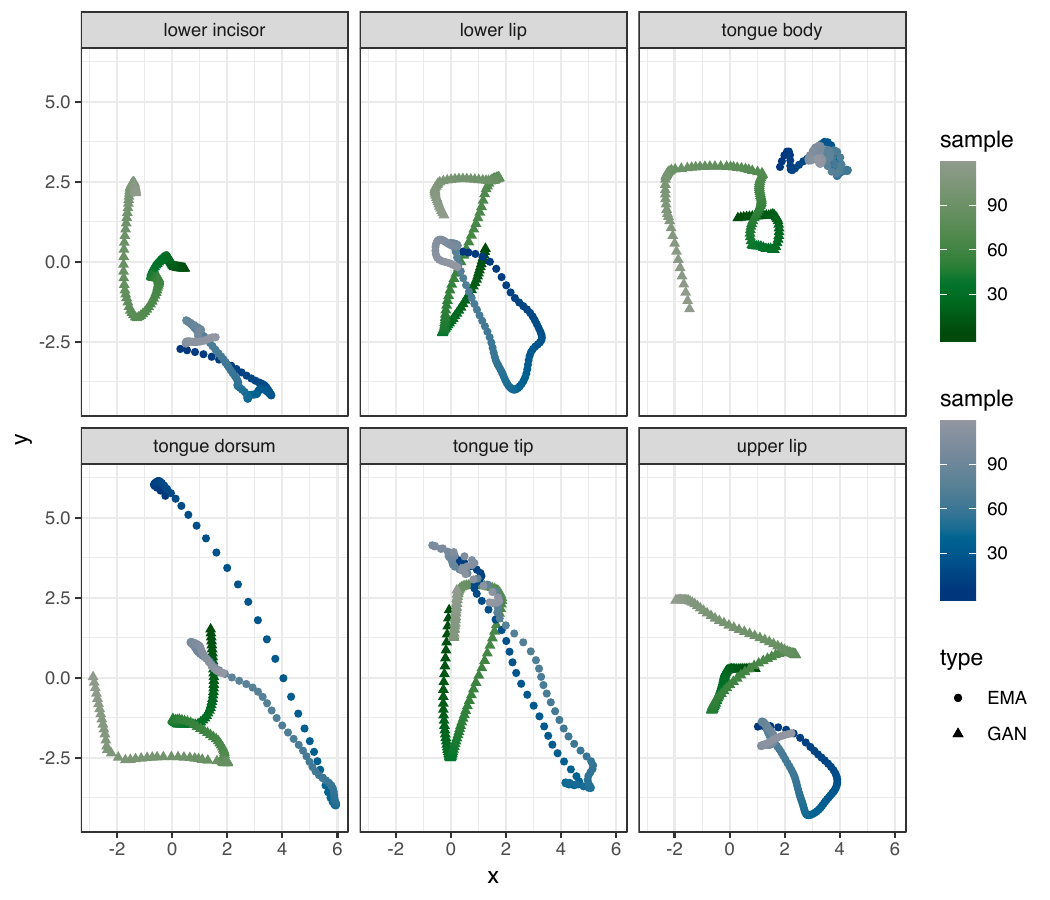}
    
    \caption{Real EMA channels (blue circles) and smoothed, generated EMA (green triangles) in 2D space for output transcribed as \textit{wash} (top) and \textit{fast} (bottom). Real EMA is multiplied by 3.0 for comparison. Temporal dimension (sample) is represented with shading. Note the similarity in trajectories between generated and real channels especially for tongue tip, lower lip, and tongue body. Quantitative analysis of trajectories in Table \ref{tab:my_label} shows a high degree of correlation}
    \label{fig:fast}
\end{figure}

To interpret articulatory gestures and compare real human EMA to generated EMA, we visualize x and y-axis values in 2D space for each electrode placement.  Because the Generator has no restrictions that would penalize rapid changes (as is the case in human muscle and  movements), we smooth the generated EMA with LOESS smoothing. Figure \ref{fig:fast} contains generated and real EMA for  \textit{wash} and an innovative output \textit{fast}. Tongue tip and lower/upper lip are the most relevant articulators for \textit{wash} and \textit{fast}. We observe  very similar gestures between GAN and EMA for \textit{wash}, and an almost identical pattern the lower lip gestures for \textit{fast}.

\subsection{Quantitative comparison between generated and real EMA}
We further performed a quantative comparison between gestures for the generated and real EMA. To account for differences in timing, we perform dynamic time warping (DTW) between smoothed generated EMA and real EMA for each dimension (x and y) and each electrode placement. We then compute Pearson's product-moment correlation ($r$) on two time series data for each channel to estimate the time-aligned correlation between generated and real EMA.

\begin{table}[ht]
\centering
\begin{tabular}{lcccc}
  \hline 
  &\multicolumn{2}{c}{wash}&\multicolumn{2}{c}{fast}\\
  Place  & $x$  & $y$ &  $x$ &  $y$ \\ \hline

tongue tip  & 0.70  & 0.90 & 0.99  & 0.96 \\ 
tongue body  & 0.94  & 0.91& 0.32  & 0.79 \\ 
lower lip  & -0.52  & 0.70 & 0.85 &  0.94 \\ 
upper lip & 0.51 & 0.90& 0.64  & 0.43 \\ 
lower incisor  & 0.87 & 0.66 & 0.31  & 0.72 \\ 
tongue dorsum & 0.41  & 0.91  & 0.24  & 0.89 \\

   \hline
\end{tabular}
    \caption{ Pearson's product-moment correlation ($r$) for \textit{wash} and \textit{fast} after DTW alignment of two time series.}
    \label{tab:my_label}
\end{table}

The quantitative comparison in Table \ref{tab:my_label} reveals a high degree of correlation in gestures between real EMA and GAN-generated EMA. Tongue tip gestures in \textit{fast} are almost identical ($r=0.99$ in x-axis and  $r=0.96$ for y axis). We observe that tongue tip, lower lip, and tongue body have highest correlations and the y-axis is better correlated than the x-axis. This is expected as vertical movements are more consequential in these words.

\subsection{Limitations \& future directions}

Despite the training complexities discussed in Section \ref{intro}, the intelligibility of ArticulationGAN's outputs is not substantially lower than that of WaveGAN (Table \ref{tab:intelligible}).  ArticulationGAN outputs a higher proportion of innovative intelligible outputs. This is not unexpected from cognitive modeling perspective: speech production (articulatory learning) is substantially more difficult than speech perception (acoustic learning), and innovative outputs are common during articulatory learning.

EMA data is a very low-dimensional representation of articulation in human speech. Adding articulatory representations (e.g.~more channels or additional articulation data types) might improve performance and provide higher resolution insights about articulation. Also, our model operates with a single Discriminator and a single 5-layer Generator that needs to generate 13 1D channels, which may impact performance. Adding multiple subdiscriminators has also been shown to increase performance in the GAN framework \cite{kong20}.

\section{Conclusion}

This paper proposes a new model for unsupervised learning of articulatory gestures in human speech production. To our knowledge, we present the first fully unsupervised deep generative network that learns to generate articulatory representation from latent noise based exclusively on the audio inputs. We argue that the Articulatory Generator learns to generate human-like articulatory representations and propose a technique to quantitatively estimate the similarities.

\clearpage

\bibliographystyle{IEEEbib}
\bibliography{bibliography.bib}

\end{document}